\documentclass[apj]{emulateapj}
\usepackage{amsmath}
\usepackage{amssymb}
\usepackage{euscript}
\usepackage{algorithm2e,algorithmic}
\usepackage{bbold}
\usepackage{paralist}

\def\gtorder{\mathrel{\raise.3ex\hbox{$>$}\mkern-14mu
             \lower0.6ex\hbox{$\sim$}}}
\def\ltorder{\mathrel{\raise.3ex\hbox{$<$}\mkern-14mu
             \lower0.6ex\hbox{$\sim$}}}

\slugcomment{Submitted to ApJ}

\shortauthors{Ofek \& Zackay}

\begin{document}




\title{Optimal matched filter in the low-number count Poisson noise regime and implications for X-ray source detection}
\author{Eran O. Ofek\altaffilmark{1} \& Barak Zackay\altaffilmark{1}}
\altaffiltext{1}{Benoziyo Center for Astrophysics, Weizmann Institute
  of Science, 76100 Rehovot, Israel}

\begin{abstract}

Detection of templates (e.g., sources) embedded in low-number count
Poisson noise is a common problem in astrophysics.
Examples include source detection in X-ray images, $\gamma$-rays, UV,
neutrinos, and search for clusters of galaxies and stellar streams.
However, the solutions in the X-ray-related literature are sub-optimal --
in some cases by considerable factors.
Using the lemma of Neyman-Pearson we derive the optimal statistics for template
detection in the presence of Poisson noise.
We demonstrate that this method provides 
higher completeness, for a fixed false-alarm probability value,
compared with filtering the image with the point-spread function (PSF).
In turn, we find that filtering by the PSF is better than filtering
the image using the Mexican-hat wavelet (used by {\tt wavedetect}).
For some background levels, our method improves the sensitivity of source
detection by more than a factor of two over the popular Mexican-hat wavelet
filtering.
This filtering technique can also be used also for fast
PSF photometry and flare detection, and it is efficient, as well as
straight forward to implement.
We provide an implementation in {\tt MATLAB}.

\end{abstract}


\section{Introduction}
\label{sec:Introduction}

The detection of a signal for which the template is roughly known,
in the presence
of Poisson noise, is considered to be a notorious problem.
In astronomy, this problem appears regularly in the context
of source detection in low-number count imaging like X-ray,
UV, $\gamma$-ray and neutrino detectors,
as well as detection of clustering of objects
(e.g., cluster of galaxies identification).

Given the importance of this problem for X-ray waveband astronomy,
several techniques for source detection, in the presence of Poisson
noise, were developed. However, as we show here,
previously available solutions are sub-optimal\footnote{By sub-optimal we mean that information is being lost.
One of the consequences of sub-optimality is that for a fixed false-alarm
probability the source-finding completeness is lower
compared with the optimal solution.}.

In the presence of additive white Gaussian noise,
it is well known that the best way to find a signal with a
roughly known template,
is to cross-correlate the data with the expected signal template.
This operation is called matched filtering, and it stands at the
base of many detection algorithms such as the LIGO gravitational wave detection
(e.g., Owen \& Sathyaprakash 1999),
source extraction in astronomical images
(e.g., Bertin \& Arnouts 1996) and many more applications in
virtually any field of science and engineering.
This simple, but powerful result is the consequence of the Neyman-Pearson
lemma (Neyman \& Pearson 1933).
The derivation of the Gaussian-noise
matched filter from the lemma of Neyman Pearson (e.g., \S\ref{sec:Algo}) proves
that the matched filter is the best solution and not only the best linear solution.

The two most common source detection techniques in X-ray astronomy
are {\tt wavedetect} (Freeman et al. 2002) and the sliding cell method (Harnden et al. 1984).
Both methods are in fact filtering techniques, while {\tt wavedetect}
filters (i.e., cross-correlates) the image
with a wavelet of some form, in the sliding-cell method we filter the image with a top-hat function.
In these previous algorithms,
the filter (e.g., wavelet) was selected without optimality proof or rigorous justification.

Here, we use the lemma of Neyman-Pearson
to derive the optimal source detection technique for the case of Poisson noise.
Not surprisingly, when the count-rate is high, this result converges to
the well known Gaussian-noise matched filter.
We also discuss extensions of this algorithm to flare detection
and point-spread function (PSF) photometry.
We demonstrate that this new technique provides better results than previously used methods,
and we provide an implementation in {\tt MATLAB}.

The structure of this paper is as follows.
In \S\ref{sec:Algo} we derive the new algorithm,
and in \S\ref{sec:ext} we present extensions of this method.
In \S\ref{sec:Im} we discuss some implementation considerations.
A step-by-step description of the algorithm is outlined in \S\ref{sec:step}, 
while in \S\ref{sec:sim} we test and compare the algorithm with other methods using simulations.
We briefly present our code in \S\ref{sec:code}, and we conclude in \S\ref{sec:Disc}.

\section{Derivation of the Poisson-noise matched filter}
\label{sec:Algo}

Our goal is to derive the optimal template-detection method when the
noise is Poisson.
Our derivation applies for a signal with any number of dimensions.
The starting point is the lemma of Neyman-Pearson that states:
When performing a hypothesis test between two
simple\footnote{Simple hypothesis specifies the population distribution completely. In such an hypothesis, there are no free parameters.} hypotheses
$\mathcal{H}_{0}$: $\theta=\theta_{0}$ and $\mathcal{H}_{1}$: $\theta=\theta_{1}$,
the likelihood-ratio test which rejects $\mathcal{H}_{0}$
in favor of $\mathcal{H}_{1}$ when
\begin{equation}
\Lambda(x) = \frac{\mathcal{L}(x|\mathcal{H}_{0}) }{ \mathcal{L}(x|\mathcal{H}_{1} )}\le\eta,
\label{eq:LRT}
\end{equation}
where the probability $\mathcal{P}(\Lambda(x)\le\eta|\mathcal{H}_{0})=\alpha$,
is the most powerful\footnote{The power of a hypothesis test is the
probability that the test correctly rejects the
null hypothesis ($\mathcal{H}_{0}$) when the
alternative hypothesis ($\mathcal{H}_{1}$) is true.}
test at significance level $\alpha$ for a threshold $\eta$.
If a test is most powerful for all $\theta_{1}\in \Theta_{1}$, it is said to be uniformly most powerful (UMP)
for alternatives in the set $\Theta_{1}$.
Here $\mathcal{L}$ is the likelihood function.
In some cases, it is possible to use tools such as
the Karlin-Rubin theorem (e.g., Casella \& Berger 2008) to derive
optimal solutions even for non-simple
hypotheses\footnote{For example, in source detection where the noise is additive white gaussian noise, the matched filter solution is optimal regardless of the flux of the source, which is a free parameter.}.
To summarize, the lemma of Neyman-Pearson provides
the methodology to construct the optimal algorithm
for simple hypothesis testing.

We denote the measured data by $M$.
For example, $M$ could be a two-dimensional image in
raw counts (e.g., number of photons, electrons or events),
in which we
would like to find sources with some known point spread function (PSF)
or template denoted by $P$.
We assume that the pixels in $M$ are independent Poisson random variables.
Given the data $M$, we would like to make a decision between two hypotheses.
The null hypothesis ($\mathcal{H}_{0}$) that there is no source
at position $q_{0}$ in the data, and the alternative hypothesis ($\mathcal{H}_{1}$)
that there is a source at position $q_{0}$ in the data.
The measurements ($M$), in our case, are drawn from the Poisson distribution.
The Poisson probability density function to get the measurement $M$
given the expectancy $\lambda$ is
\begin{equation}
\mathcal{P}_{\rm poi}(M|\lambda) = \frac{\lambda^{M} e^{-\lambda}}{M!},
\label{eq:Poi}
\end{equation}
The model for the null hypothesis is:
\begin{equation}
\mathcal{H}_{0} : M(q) = poi(B),
\label{eq:H0}
\end{equation}
and for the alternative hypothesis is:
\begin{equation}
\mathcal{H}_{1}(q_{0},F) : M(q) = poi(B + F P(q-q_{0})).
\label{eq:H1}
\end{equation}
Here $poi(E)$ is a Poisson random variable with expectancy $E$,
$M(q)$ is the measured data with multi-dimensional coordinate $q$
(e.g., $x,y$ coordinates for an image), $B$ is the expectancy value for the background level
(e.g., as estimated from a region around the $q_{0}$ position),
$F$ is the unknown flux of the source we would like to detect,
and $P(q)$ is the PSF (or more generally a template)
we would like to search in our data.
The PSF is normalized to unity 
\begin{equation}
\int_{q}{P(q)dq}=1.
\label{eq:Norm}
\end{equation}
However, $\mathcal{H}_{1}$ is a composite hypothesis
as it has a free parameter, $F$.
This will be treated later and for the time
being we will assume that $F$ is known.

Following the directive of the Neyman-Pearson lemma, we
write the $\log$-likelihood difference at position $q_{0}$
\begin{eqnarray}
-\ln{\Lambda(q_{0})} & = -\sum_{q}{\{ M(q)\ln{B} - B - \ln{M(q)!} \} } \cr
 & + \sum_{q}{\{ M(q)\ln{[B+FP(q-q_{0})]} \} } \cr
 & - \sum_{q}{\{ B + FP(q-q_{0}) + \ln{M(q)!} \} }.
\label{eq:lnLambda}
\end{eqnarray}
Some of the terms in this expression cancel out.
Furthermore, we are allowed to remove any term that does~not depend on the data
(i.e., $FP$) -- such terms contribute a fixed constant which we can absorb
into the threshold ($\eta$).
Denoting the log-likelihood difference by $S$
and simplifying, we get
\begin{equation}
S(q_{0})\equiv-\ln{\Lambda(q_{0})} = \sum_{q}{M(q)\ln{[1+ \frac{F}{B}P(q-q_{0})]}}.
\label{eq:lnLambda2}
\end{equation}
Equation~\ref{eq:lnLambda2} can be identified with the cross-correlation
operation and therefore can be re-written,
simultanously for all positions in the image, as
\begin{equation}
S = M \otimes \overleftarrow{P_{\rm poi}},
\label{eq:lnLambdaConv}
\end{equation}
where $\otimes$ denotes convolution, $\overleftarrow{}$ denotes coordinates reversal (e.g., $x,y\rightarrow -x,-y$),
and $P_{\rm poi}$ is the Poisson-noise optimal filter
\begin{equation}
P_{\rm poi}=\ln{(1+ \frac{F}{B}P)}.
\label{eq:Ppoi}
\end{equation}
This can also be written in Fourier space as
\begin{equation}
\widehat{S} = \widehat{M} \overline{\widehat{P_{\rm poi}}}.
\label{eq:lnLambdaF}
\end{equation}
Here the $\widehat{}$ symbol denote the Fourier transform,
the bar symbol denotes the complex conjugation.
The fact that the bar sign is over the hat symbol means that the complex conjugate
operation follows the Fourier transform.

In the language of matched filtering we can identify $\ln{(1+FP/B)}$
as the filter. 
We note that our convention is to subtract the background
expectation value from the image prior to filtering,
and not to renormalize $P_{\rm poi}$ to unity.
Other conventions are valid as long as they are applied consistently.

It is interesting to note that Taylor expansion of Equation~\ref{eq:lnLambda2} leads
to the well known matched filter in the case of Gaussian noise
(i.e., in this case the filter is $P$).

However, we are not done yet.
The flux of the source we would like to find ($F$)
is unknown apriori, and Equation~\ref{eq:lnLambdaF} is non-linear in $F$.
More generally, $\mathcal{H}_{1}$ 
is not a simple hypothesis as it has a free parameter $F$.
Moreover, in this case, it is possible to show that there is no value of $F$
which is uniformly most powerful (see Appendix~\ref{Ap:Optimal}).
A solution for this problem is to set a flux threshold
$F_{\rm th}$ and to filter the image
with several filters, each with a different $F$ value ($F\ge F_{\rm th}$).
However, as we demonstrate in Appendix~\ref{Ap:Optimal}
for any practical application it is enough to filter
the image with a single value of $F=F_{\rm th}$.
Although this is not uniformly most powerful,
we show that the loss of sensitivity due to this approximation
is of the order of $\ltorder1$\%.
Therefore, for detection purposes we have to set $F$ in
Equation~\ref{eq:lnLambdaF} to our preferred flux threshold.

However, there are two additional complications:
(i) We do~not know apriori what is the flux threshold
associated with our desired false-alarm probability;
(ii) In order to find sources in $S$ we need  to look
for local maxima in $S$ and calculate the probability
to get this, or larger, local maxima value given the
probability distribution of values in $S$.
However, $S$ is the result of summation
of many Poisson random variables with weights, and the resulting
probability distribution is complicated.

These problems can be solved numerically using the following scheme:
Given the background ($B$) and PSF ($P$), and a desired false alarm probability $\beta$,
we would like to find a self-consistent flux threshold
($F_{\rm th}$) such that
\begin{equation}
\int_{S_{\rm th}}^{\infty}{ \mathcal{P}_{\rm S}(S)dS} = \beta,
\label{eq:selfcon}
\end{equation}
where
\begin{equation}
S_{\rm th} = (F_{\rm th}-B)S_{\rm F},
\label{eq:Sth}
\end{equation}
and $S_{\rm F}$ is the normalization factor that transforms the units of the score statistics $S$
to flux-count units $F$.
Note that $F_{\rm th}$ has the flux units of the measured data $M$
prior to background subtraction, while $S_{\rm F}$ is measured
in the background subtracted $S$ image.
This normalization is simply the summation over a PSF with unit flux
multiplied by its filter:
\begin{equation}
S_{\rm F} = \sum_{q}{P(q)\ln{(1+F_{\rm th}P(q)/B)}}.
\label{eq:SF}
\end{equation}
Finally, $\mathcal{P}_{\rm S}$ is the probability distribution function of the background
pixels in $S$.
We note that in Equation~\ref{eq:Sth} we subtract $B$ from $F_{\rm th}$
as our convention is to subtract the background from the images.
While $F_{\rm th}$ refers to the flux in the original image,
$S_{\rm th}$ refers to the flux in the background subtracted $S$ image.

A simple method to find $\mathcal{P}_{\rm S}$ is using numerical
Monte-Carlo simulations.
In such simulations we need to generate Poisson random images
with expectancy $B$, subtract the expectancy $B$, and
filter them with our matched filter
\begin{equation}
S_{\rm sim} = (poissrnd_{m\times n}(B) - B)\otimes\overleftarrow{P_{\rm poi}},
\label{eq:Ssim}
\end{equation}
where $poissrnd_{m\times n}(E)$ is a function that generates Poisson random variables with
expectancy $E$ over an image of size $m\times n$.
Now, $\mathcal{P}_{\rm S}$ is given by the probability distribution (histogram) of values in $S_{\rm sim}$.
Finally, a good first-iteration guess for $F_{\rm th,0}$ is given
implicitly\footnote{Can be calculated using, e.g., {\tt poissinv} in MATLAB.} by
\begin{equation}
1-\beta = \int_{0}^{F_{{\rm th}, 0}}{\mathcal{P}_{\rm poi}(f|4\pi \sigma^{2} B)df  },
\label{eq:Fth0}
\end{equation}
where $\sigma$ is the width of the PSF (e.g., $\sigma$ of a Gaussian PSF).
In practice we find that only a few iterations are required in order to find
the self-consistent $F_{\rm th}$ and $S_{\rm th}$.

An interesting question is by how much this method improves over a naive
matched filter (i.e., cross correlating the image with its PSF)?
or other filters, like the one advocated by the {\tt wavedetect}
method (Freeman et al. 2002)?
We address these questions, using simulations, in \S\ref{sec:comp}.

Finally, we would like to emphasize that other filters
can be designed such that they will have the same false-alarm probability ($\beta$).
However, detection based on such filters will always have
lower completeness (i.e., higher $F_{\rm th}$)
compared with the method presented in this paper --
this is a consequence of the Neyman-Pearson lemma.

\section{Additional applications}
\label{sec:ext}

\subsection{PSF photometry}

The statistics $S$ can be used to perform PSF photometry
in the case of Poisson noise.
The prescription is similar to the one suggested in Zackay \& Ofek (2017a)
for the Gaussian noise case.
In a nutshell, each pixel in S contains the PSF-weighted sum
of its neighboring pixels.
Therefore, providing a PSF photometry estimator.

In order to measure the flux of a source one needs to convert $S$
to flux units. The flux estimator, $\widetilde{F}$
is given by dividing $S$ by $S_{\rm F}$ (Eq.~\ref{eq:SF})
\begin{equation}
\widetilde{F} = \frac{S}{\sum_{q}{P(q)\ln{(1+\widetilde{F}P(q)/B )} }}.
\label{eq:FluxM}
\end{equation}
The estimator for the flux of the source is simply $\widetilde{F}$ at the position
of the source.
However, this equation is not implicit and therefore
we need to solve it iteratively for $\widetilde{F}$.
Alternatively, a good approximation is to calculate this equation for several 
$F$ values, on the right-hand side, in some logarithmic steps.
Another minor problem with this approach is that it
ignores the exact (sub-pixel) position of the source
(see additional discussion and details in Zackay \& Ofek 2017a).

\subsection{Flare detection}
\label{sec:flare}

We note that the method presented in this paper can be extended
to the problem of flare detection in the Poisson-noise regime
(e.g., Scargle et al. 2013).
In this case all we need to do is to match filter the image
also in the time dimension, not only in the position dimensions.
Since the position (PSF) and time (flare template) dimensions
are independent, the expression for the filter is simply:
\begin{equation}
P_{poi,flare}(x,y,t) = \ln{(1+\frac{F}{\dot{B}}PT_{flare})}.
\label{eq:Pflare}
\end{equation}
Here $\dot{B}$ is the background per unit time,
and $T_{flare}$ is the flare template as a function of time
(counts per unit time normalized to unity).
Note that in this case the cross-correlation operation is
performed in both position and time.

This simple extension allows us to filter the image
in both the temporal and position dimensions, and to optimally
detect flares which have some known template.
Even if the flare template is unknown, one may use
a bank of top-hat or fast-rise exponential decay functions
with various time scales.
The main advantage over other methods (e.g., Scargle et al. 2013)
is that it will have increased sensitivity due to the additional filtering
in the position dimensions.
We may further discuss and demonstrate this idea in a future publication.

\section{Implementation considerations}
\label{sec:Im}

\subsection{Flat fielding / exposure maps}

In some cases, our photon-count image has to be flat fielded.
A simple solution is to use the flat field or exposure map to estimate the background $B$ locally,
and to apply our filter locally with the appropriate background.
This is reasonable as long as the background varies slowly compared to the PSF size.

\subsection{Background estimation}
\label{sec:Back}

It is straight forward to estimate the expectancy of a Poisson random variable, by taking the mean value of the counts.
However, any decent background estimator should ignore regions that contains
bright sources or artifacts.
We suggest that a good way to estimate the background
is to do this iteratively. I.e., detect sources, estimate the background in
source-free regions, and run the source detection again.

A less robust, but somewhat simpler approach, that may work in some cases,
is to ignore pixels with $>1$ events and to use the ratio between the number of pixels with 1 event and 0 events as a robust estimator for the background:
\begin{equation}
\frac{\mathcal{P}_{\rm poi}(1|B)}{\mathcal{P}_{\rm poi}(0|B)} = \frac{B e^{-B}}{e^{-B}} = B.
\label{eq:BackEst}
\end{equation}

\subsection{PSF}

It is important to emphasize that a perfect knowledge of the PSF is not required.
This point is further demonstrated in \S\ref{sec:comp}
where we show that small differences between the optimal filter of the image
and the used filter have negligible effect on the source-detection results.
This indicates that a perfect knowledge of the PSF is not a requirement
(see also a discussion regarding the Gaussian-noise case in
Zackay \& Ofek 2017a).

\subsection{PSF variations}

In X-ray images the PSF may vary substantially across the image.
A way to deal with this problem is to partition the image to regions
in which the PSF (and background) are roughly constant and apply the
filter for each such sub image separately.
We note that it is possible to code an efficient convolution
algorithm that allows the PSF to vary smoothly across the image.

\subsection{Energy dependent PSF and background}

In many cases in high-energy astronomy, the instrument
is sensitive to a wide range of energies, and the PSF
and background are energy dependent.
Extending our optimal filter to an energy dependent PSF
and background is straight forward.
One needs to break the image into multiple energy channels, and filter each
energy-channel image with its own filter.
This approach is valid as long as the energy
range in which the PSF and background change is larger than
the uncertainty in the photons energy.
The result is a score image per energy channel ($S_{\rm E}$).
Finally, since the energy channels provide independent information,
the optimal statistics for source detection is given by the
summation over all $S_{\rm E}$
\begin{equation}
S = \sum_{\rm E}{S_{\rm E}}.
\label{eq:SES}
\end{equation}
We note that in this approach one can also account for
the predicted spectrum of the source.
In this case $F_{\rm th}$ can be derived based
on simulations that take into account
the expected value of the background, PSF, and source
flux, as a function of energy.

%

%

\section{Step by step algorithm}
\label{sec:step}

Our Poisson-noise filtering algorithm
can be summarized by the following steps:
For each image, or section of the image, in which the
properties of the PSF and background are uniform:
\begin{enumerate}
   \item Estimate the local background ($B$; see \S\ref{sec:Back}).
   \item Subtract the background expectancy from the image.
   \item Estimate or use the known PSF ($P$).
   \item Select a false alarm probability $\beta$.
   \item Given $\beta$ solve Equations~\ref{eq:selfcon},~\ref{eq:Sth}, and~\ref{eq:SF} for $F_{\rm th}$.\\
         This can be done either by interpolating Table~\ref{tab:Fth},
         or direct calculation using the following steps:\\
         (i). Set $F_{\rm th}$ to the guess value using Equation~\ref{eq:Fth0}.\\
         (ii). Simulate a background image ($S_{\rm sim}$), subtract the background expectancy,
            and filter (Equation~\ref{eq:Ssim}).\\
         (iii). Calculate $S_{\rm th}$, which is given by the $\beta$ upper quantile of the values
            in $S_{\rm sim}$.\\
         (iv). Calculate the flux normalization $S_{\rm F}$ (Equation~\ref{eq:SF}).\\
         (v). Set $F_{\rm th}=S_{\rm th}/S_{\rm F}+B$, and go to step (ii),
            until convergence.\\
         (vi). In the final iteration you have $F_{\rm th}$ and $S_{\rm th}$.\\
   \item Calculate the Poisson-noise filter (Equation~\ref{eq:Ppoi}) by setting $F=F_{\rm th}$
         from step 5.
   \item Filter the image. This can be done in real space (Equation~\ref{eq:lnLambdaConv})
         or using FFT (Equation~\ref{eq:lnLambdaF}).
         If using FFT, make sure to FFT shift the filter before the cross correlation operation, such that the filter peak will be at the image origin (corners).
         Otherwise the convolution operation will introduce a shift to the result.
   \item Find local maxima in the filtered image $S$.
         If the value of the local maxima is larger than $S_{\rm th}$
         then declare a source detection in the pixel of the local maxima.
\end{enumerate}

We note that the threshold should be adjusted to the look-elsewhere effect --
the number of trials per image is $\approx N_{\rm pix}/(\sigma^2)$,
where $N_{\rm pix}$ is the number of pixels in the image and $\sigma$
is the Gaussian PSF $\sigma$ (in pixels).

Furthermore, we note again that our convention is
to normalize P to unity, but not to renormalize $P_{\rm poi}$ to unity,
and to subtract the background expectation value from the image prior
to filtering.
Other conventions are valid as long as they are applied consistently.

\section{Simulations}
\label{sec:sim}

In \S\ref{sec:pS} we discuss the properties of the $\mathcal{P}_{\rm S}$ probability
density function based on simulations, and in \S\ref{sec:tab} we present tabulated values
of the detection threshold as a function of the background, PSF width,
and the false-alarm probability.
In \S\ref{sec:comp} we compare the Poisson Matched Filter with
filtering with the PSF and the Mexican-hat (Ricker) wavelet.

\subsection{The properties of $\mathcal{P}_{\rm S}$}
\label{sec:pS}

\begin{figure}
\centerline{\includegraphics[width=8cm]{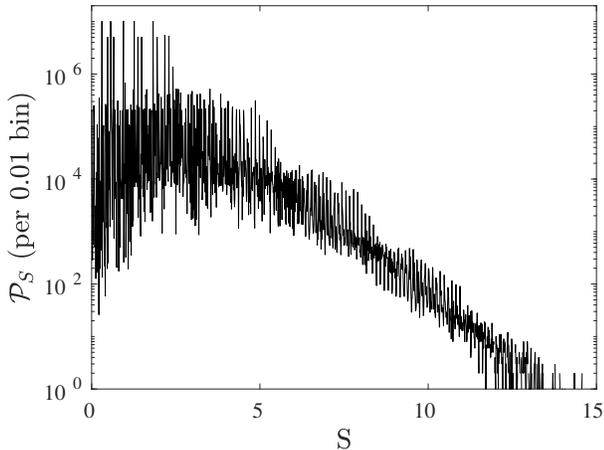}}
\caption{Unnormalized $\mathcal{P}_{\rm S}$ per 0.01 count bins, for the case of
$B=0.005$\,counts\,pix$^{-1}$, Gaussian PSF with $\sigma=2$\,pix and $\beta=10^{-3}$.
The highest bin contains over $10^{7}$ events so the precision of this plot
around the maximum is better than $10^{-3.5}$.
\label{fig:p_S_simulation_sigma2_beta0.001_B0.005}}
\end{figure}

In general $\mathcal{P}_{\rm S}$ is a complicated distribution.
Inspection of
a simulated $\mathcal{P}_{\rm S}$ shows that it is not a smooth function, but instead
looks like peaks and dips over a smooth envelope.
This behavior is expected, as $S$ originates from a linear
combination of discrete Poisson distributions.
To emphasize this point, Figure~\ref{fig:p_S_simulation_sigma2_beta0.001_B0.005} shows
$\mathcal{P}_{\rm S}$ for $B=0.005$\,counts\,pix$^{-1}$, Gaussian PSF with $\sigma=2$\,pix and $\beta=10^{-3}$.
Although, this is a complicated function, it can be calculated.

\begin{figure}
\centerline{\includegraphics[width=8cm]{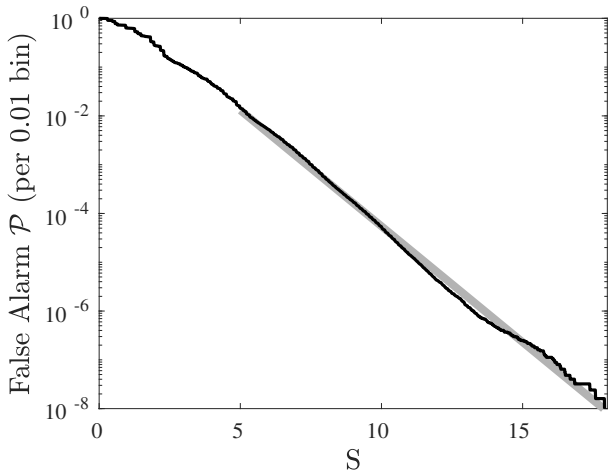}}
\caption{The false alarm probability, as a function of $S$ per 0.01 bin, for the case of
$B=0.005$\,counts\,pix$^{-1}$, Gaussian PSF with $\sigma=2$\,pix and $\beta=10^{-3}$.
The gray line shows the best fit power law, which has a power-law index
of $\cong-1/2$.
\label{fig:Cummul_p_S_simulation_sigma2_beta0.001_B0.005}}
\end{figure}

A problem with our approach for calculating $\mathcal{P}_{\rm S}$ is that
it is not an efficient method for calculating the threshold for
very small $\beta$, as in this case a very large number
of simulations is required.
In Figure~\ref{fig:Cummul_p_S_simulation_sigma2_beta0.001_B0.005} we show the false alarm probability as a function of $S$,
where the false alarm probability is defined as $\int_{S_{\rm th}}^{\infty}{\mathcal{P}_{\rm S}dS}$.
This plot suggests that the false alarm probability can be approximated
(to within a factor of 2, for large values of $S$) as a power law with slope of about $-1/2$.

An alternative analytical
method for calculating $\mathcal{P}_{S}$ (almost) exactly and without the need
for a large number of simulations is to sum the probability over all possible
combinations of photons in pixels.
If the background is low (e.g., $B\ll1$) and the PSF is circularly symmetric,
then the number of possible permutations
(e.g., all possibilities, except those with negligible probability,
of the number of counts in each pixel in the PSF)
is not very large and can be computed by finite number of summations.

\subsection{Tabulated flux thresholds}
\label{sec:tab}

In \S\ref{sec:code}, we provide tools for calculating $F_{\rm th}$
and $S_{\rm th}$ as a function of the PSF, $B$, and $\beta$.
However, sometimes it is useful (or more efficient) to
derive these thresholds from interpolation of a pre-calculated grid.

In Table~\ref{tab:Fth} we list $F_{\rm th}$ and $S_{\rm th}$, for a Gaussian PSF,
as a function of the PSF width $\sigma$, the background $B$,
and the false alarm probability $\beta$.
We note that $S_{\rm th}$ can be calculated from $F_{\rm th}$,
but here we list both for convenience.
The table is also available electronically\footnote{http://weizmann.ac.il/home/eofek/matlab/doc/Poisson\_Matched\_Filter.html}.

\begin{deluxetable}{lllll}
\tablecolumns{5}
\tablewidth{0pt}
\tablecaption{Flux threshold as a function of $B$, $\sigma$, and $\beta$}
\tablehead{
\colhead{B}               &
\colhead{$\sigma$}        &
\colhead{$\beta$}         &
\colhead{$F_{\rm th}$}      &
\colhead{$S_{\rm th}$}      \\
\colhead{(counts\,pix$^{-1}$)}        &
\colhead{(pix)}           &
\colhead{()}              &
\colhead{(counts)}        &
\colhead{()}      
}
\startdata
 0.0001 &  1.000 & 0.00010  &        1.14  &    7.51\\
 0.0001 &  1.000 & 0.00032  &        1.06  &    6.94\\
 0.0001 &  1.000 & 0.00100  &        0.83  &    5.20\\
 0.0001 &  1.468 & 0.00010  &        1.16  &    6.74\\
 0.0001 &  1.468 & 0.00032  &        1.12  &    6.48
\enddata
\tablecomments{$F_{\rm th}$ and $S_{\rm th}$, for a Gaussian PSF,
as a function of the PSF width $\sigma$, the background $B$,
and the false alarm probability $\beta$.
Here we present the first few lines. The full table is available in the electronic version of the paper.
In addition it is available online (\S\ref{sec:code}).}
\label{tab:Fth}
\end{deluxetable}

\subsection{Comparison with wavelets and PSF filtering}
\label{sec:comp}

\begin{figure}
\centerline{\includegraphics[width=8cm]{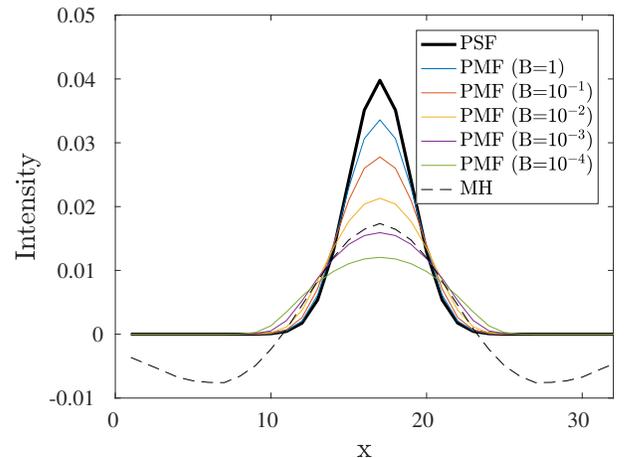}}
\caption{A 1-D cut through a Gaussian PSF (solid-bold line), Mexican-hat wavelet (dashed line),
and the normalized Poisson-noise matched filters
(PMF; i.e., Equation~\ref{eq:Ppoi}), for various background levels
(indicated in legend).
The order of the PMFs in the legend corresponds to their maximum height.
In all cases we assumed Gaussian PSF with width $\sigma=2$ and a false alarm probability of $\beta=10^{-3}$.
\label{fig:Filters_PSF_vs_PMF}}
\end{figure}

Figure~\ref{fig:Filters_PSF_vs_PMF} shows a 1-D cut through a Gaussian PSF,
the Mexican-hat wavelet, and
the normalized Poisson filters (i.e., Equation~\ref{eq:Ppoi}), 
for various background levels.
In all cases we assumed a Gaussian PSF with width $\sigma=2$ and a false alarm probability of $\beta=10^{-3}$.
For the Mexican hat we selected the support (width) that roughly maximizes the completeness (see below).
From this figure it is apparent that the Poisson-noise matched filter converges to the PSF when the background is high.
In fact this plot suggests that when the background is larger than a few
counts per PSF area, the
Gaussian-noise matched filter (i.e., the PSF) is an excellent approximation to the Poisson-noise optimal filter.
Next we see that for low background level the Mexican hat, Gaussian-noise, and Poisson-noise matched filters
are very different. Since we know by construction that the Poisson-noise matched filter is
optimal, we are led to suspect that using the Gaussian-noise or Mexican-hat
filters leads to a loss of information.
In other words, we expect that if we will use a sub-optimal filter (e.g., PSF or wavelet), for a fixed
false alarm probability flux threshold, we will have lower completeness compared with the optimal filter.

In order to compare the various methods we calculate, using simulations,
the detection completeness as a function of the source flux
when the false-alarm probability is fixed.
We present two sets of simulations. In the first simulation we used
$B=0.005$\,counts\,pix$^{-1}$, a Gaussian PSF with
width of $\sigma=2$\,pix, and false alarm probability $\beta=10^{-3}$,
while in the second simulation we used $B=0.1$\,counts\,pix$^{-1}$,
and the other parameters remained the same.
For each filter, we translated the false-alarm probability to $S_{\rm th}$.
Then we simulated $10^{6}$ small images,
each with a single source that follows the Gaussian PSF, with a known flux.
We cross-correlated these images with each filter we test and check if the
source position in the filtered images is above the detection threshold.
These simulations were used to estimate the source-detection completeness,
as a function of flux.

\begin{figure}
\centerline{\includegraphics[width=8cm]{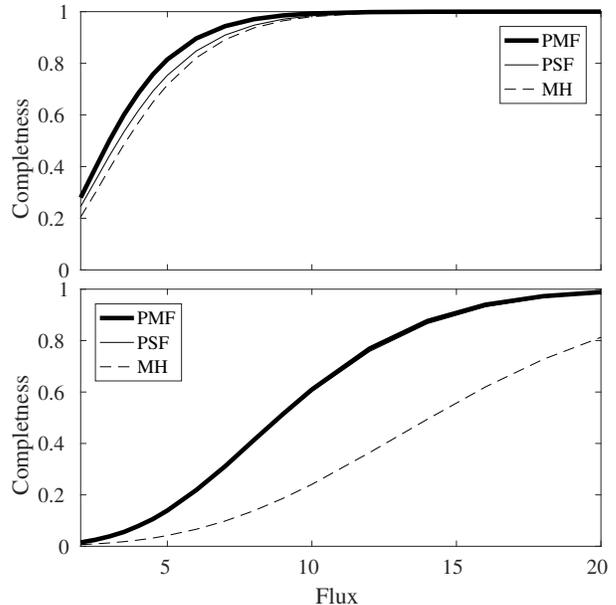}}
\caption{The source detection completeness, as a function of source flux in counts, using
Poisson-noise Matched Filter (PMF; solid-bold line), PSF filtering (solid-thin line), and the Mexican-hat (MH) filter (dashed line).
The upper plot is for $B=0.005$\,counts\,pix$^{-1}$, a Gaussian PSF with
width of $\sigma=2$\,pix, and false alarm probability $\beta=10^{-3}$.
We note that for this case $F_{\rm th}=2.73$\,counts.
The lower plot is the same but for $B=0.1$\,counts\,pix$^{-1}$.
For this case $F_{\rm th}=8.4$\,counts.
We note that in the lower panel, the Gaussian-noise filter line
is slightly below the PMF line,
and therefore the two lines are hardly separable.
\label{fig:Completness}}
\end{figure}

\begin{figure}
\centerline{\includegraphics[width=8cm]{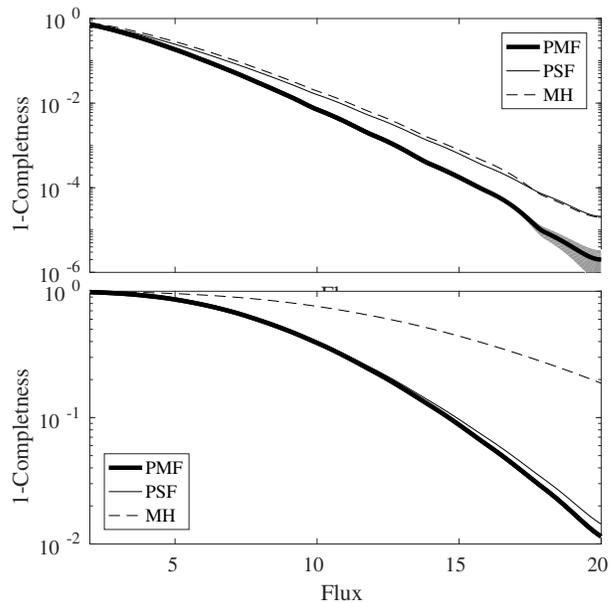}}
\caption{Like Fig.~\ref{fig:Completness}, but for the one minus the source detection completeness, as a function of source flux in counts, using
Poisson-noise Matched Filter (PMF), PSF filtering, and the Mexican-hat (MH) filter.
The gray zone around the PMF filter represents the 1-$\sigma$ uncertainty in the simulations.
The upper plot is for $B=0.005$\,counts\,pix$^{-1}$, a Gaussian PSF with
width of $\sigma=2$\,pix, and false alarm probability $\beta=10^{-3}$.
The lower plot is the same but for $B=0.1$\,counts\,pix$^{-1}$.
\label{fig:AntiCompletness}}
\end{figure}

Figure~\ref{fig:Completness} shows the completeness as a function of the source flux,
for the two simulated background levels,
for three filters: (i) the Poisson-noise filter (Equation~\ref{eq:Ppoi});
(ii) the PSF itself;
and (iii) a Mexican-hat wavelet filter.
Since with the Mexican-hat filter we are free to choose the filter width (so called support),
we used the support that maximizes the completeness.
For our $B=0.005$\,counts\,pix$^{-1}$ we find this support to be $-31.25$ to $+31.25$\,pix,
while for $B=0.1$\,counts\,pix$^{-1}$ it changed to $-22.7$ to $+22.7$\,pix.
In order to better appreciate also the completeness at high fluxes,
in Figure~\ref{fig:AntiCompletness}
we show, for the same parameters, the 1 minus completeness in logarithmic scale.

These figures demonstrate that our new filtering technique provides better results
than filtering the image by the PSF or a Mexican-hat wavelet.
For very low background levels our new method is better than PSF and wavelet filtering.
For example, for the $B=0.005$\,counts\,pix$^{-1}$ case,
at a flux of 5\,counts the completeness are 0.81, 0.75, and 0.71 for
the Poisson-noise, PSF, and Mexican-hat filters, respectively.
For 50\% completeness the flux threshold is 3.0 3.3, and 3.6 counts, respectively.
Therefore, in this case, the Poisson-noise
filter improves upon the Mexican-hat filter,
by about 20\% in depth for a fixed completeness.

For intermediate flux levels (i.e., $B=0.1$\,counts\,pix$^{-1}$;
lower panel in Figs.~\ref{fig:Completness}-\ref{fig:AntiCompletness})
the PSF filtering converges with the Poisson-noise filtering.
However, in this case the Mexican-hat (Ricker) wavelet provides poor results.
For example, for sources with flux of 10\,counts,
the Mexican-hat wavelet filtering completeness
is factor of 2.5 times lower than that of the optimal filter.
We conclude that applying our new algorithm for e.g., X-ray images, may
result in a considerable improvement in source-detection completeness
and false-alarm probability.


\section{Code}
\label{sec:code}

We developed a code for Poisson-noise optimal source detection.
The code is available as part of the
MATLAB Astronomy \& Astrophysics
Toolbox\footnote{http://weizmann.ac.il/home/eofek/matlab/} (Ofek 2014).
We supply both low level and high level functions.
Our low level functions include tools
to estimate $F_{\rm th}$ and $S_{\rm th}$ 
for arbitrary PSF based
on simulations, or for Gaussian PSF by interpolation from Table~\ref{tab:Fth};
construction of a Poisson-noise filter;
filtering; and thresholding operations.
Our high level tools include the {\tt mextractor} function
for source extraction and measurements (e.g., photometry, astrometry, shape).
This function was adapted to
support the Poisson-noise matched filter.

Given that this code will undergo improvements,
for further information we refer the reader
to the online documentation\footnote{https://webhome.weizmann.ac.il/home/eofek/matlab/doc/Poisson\_Matched\_Filter.html}.

\section{Discussion}
\label{sec:Disc}

We derive the optimal source detection algorithm in the presence
of Poisson noise. 
We demonstrate that this filter improves
the completeness, at a fixed false alarm probability,
compared with other popular techniques.
The algorithm is straightforward to implement and we provide
an implementation in {\tt MATLAB}.

The new algorithm can improve the completeness
and purity of source detection in
X-ray images, as well as UV, EUV, $\gamma$-ray,
and particle detection (e.g., TeV-photons detectors).
Furthermore, this algorithm can be adapted for other
applications such as flare detection
and detection of stars or galaxies clustering in the sky.
Following the approach of Zackay \& Ofek (2017a,b)
and Zackay, Ofek, \& Gal-Yam (2016),
the new technique can likely be extended for
image coaddition and subtraction.

\acknowledgments

E.O.O. is grateful for the support by
grants from the 
Israel Science Foundation, Minerva, Israeli ministry of Science,
the US-Israel Binational Science Foundation,
and the I-CORE Program of the Planning
and Budgeting Committee and The Israel Science Foundation.
B.Z. is grateful for the Clore fellowship foundation.

\appendix

\section{An almost optimal filter}
\label{Ap:Optimal}

\begin{figure}
\centerline{\includegraphics[width=8cm]{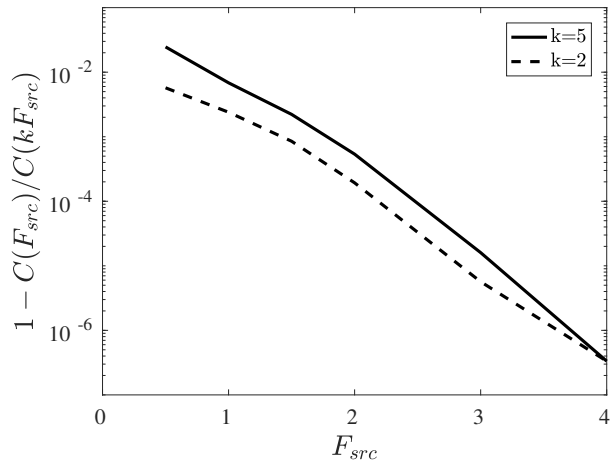}}
\caption{The relative loss in completeness due to the use of a single flux value
($F_{\rm th}$) in the Poisson-noise matched filter, as a function
of the source flux. Calculated for the case of $B=0.005$\,counts\,pix$^{-1}$.
\label{fig:NonOptimal}}
\end{figure}

A problem with our suggested filter is that it is optimal
(by construction) only when the flux of the source $F$ is known.
We note that in the Gaussian-noise case, it is possible to
show that the matched filter solution is uniformly most powerful (UMP).
However, in the Poisson noise case this is not true
(see Figure~\ref{fig:NonOptimal}).

A simple solution to this problem is to filter the image
with multiple filters (Eq.~\ref{eq:Poi}), each having a different flux ($F$) value.
Since the problem is not very sensitive to the exact shape of
the filter, there is no need to run it for many $F$ values
(e.g., several logarithmically spaced flux values).
However, in practice, using simulations we found that,
at a given false alarm probability, the
value of $F$ has small effect on the completeness.
Therefore, using a single $F$ value may be
sufficient for most applications.

\begin{figure}
\centerline{\includegraphics[width=8cm]{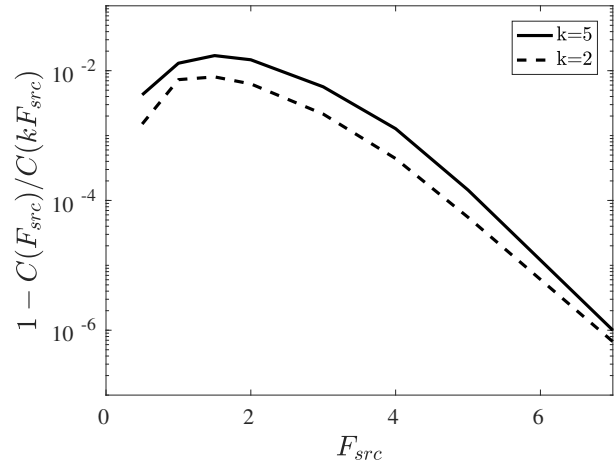}}
\caption{Like Figure~\ref{fig:NonOptimal}, but
for the case of $B=0.1$\,counts\,pix$^{-1}$.
\label{fig:NonOptimal2}}
\end{figure}

In order to test this we conducted the following simulations
with $B=0.005$\,counts\,pix$^{-1}$ and Gaussian PSF with $\sigma=2$.
For each source flux level $F_{\rm src}$ we estimated $\mathcal{P}_{\rm S}$
by applying the Poisson-noise filter (Eq.~\ref{eq:Poi}) with
$F=F_{\rm src}$. For each $F$ we calculated the value of
$S_{\rm th}$ that corresponds to
the false alarm probability of $\beta=10^{-3}$.
Next, we generated $10^{6}$ simulated images 
with a source with flux $F_{\rm src}$ and Poisson noise,
than we applied the Poisson-noise
matched filter (Eq.~\ref{eq:Poi}) with $F=kF_{\rm src}$, for $k=1,2,5$,
and calculated the detection completeness (i.e., the fraction
of simulations in which $S>S_{\rm th}$).
Figure~\ref{fig:NonOptimal} presents the $1-C(F_{\rm src})/C(kF_{\rm src})$
as a function the the source flux $F_{\rm src}$,
where $C(F)$ is the completeness at flux level $F$,
and $k=1,2,5$.
Figure~\ref{fig:NonOptimal2} shows the same but for $B=0.1$\,counts\,pix$^{-1}$.
These simulations demonstrate that in many cases,
the loss of sensitivity due to the use of the wrong $F$
in Equation~\ref{eq:Ppoi},
is below $\approx1$\%.
Therefore, we suggest that our filter with
a single $F$ value is nearly optimal,
and using a single value of $F$
is good enough for most practical applications.

\end{document}